\documentclass[aps,twocolumn,pr*,amsfonts,showpacs,superscriptaddress,longbibliography]{revtex4-2}
\DeclareUnicodeCharacter{0393}{$\Gamma$}
\usepackage{verbatim,epsfig,amsmath,amssymb,bm,epsf,graphicx,psfrag,bbold,amsthm,amsfonts}
\usepackage[bottom]{footmisc}
\usepackage{hyperref}
\usepackage{cleveref} %should be placed only after hyperref
\usepackage{enumitem}
\usepackage{framed}
\usepackage{mathrsfs}
\usepackage{esint}
\setlist{nosep}
\usepackage{placeins}
\usepackage[all]{xy}
\usepackage{color}
\usepackage[utf8]{inputenc}
\usepackage{float}
\usepackage{natbib}
\usepackage{tikz-cd}
\usepackage{verbatim}
\usepackage{leftidx}
\usepackage{physics}
\usepackage{ulem}

\usepackage{pifont}% http://ctan.org/pkg/pifont
\usepackage{braket}
\usepackage{booktabs}

\newcommand{\Loss}{L}
\newcommand{\psitheta}{\psi_{\scriptscriptstyle\theta}}
\newcommand{\psiref}{\psi_{\rm{ref}}}
\newcommand{\phitheta}{\varphi_{\scriptscriptstyle\theta}}
\newcommand{\phiref}{\varphi_{\rm{ref}}}

\newcommand{\Rb}{\mathbf{R}}
\newcommand{\rb}{{\bm r}}

\newcommand{\psiL}{\psi_{\scriptscriptstyle L}}
\newcommand{\psiMR}{\psi_{\scriptscriptstyle M\!R}}
\newcommand{\phiL}{\varphi_{\scriptscriptstyle L}}
\newcommand{\phiMR}{\varphi{\scriptscriptstyle M\!R}}
\newcommand{\psiBCS}{\psi_{\scriptscriptstyle B\!C\!S}}
\newcommand{\phiBCS}{\varphi{\scriptscriptstyle B\!C\!S}}
\newcommand{\LF}{L_{\scriptscriptstyle F}}

\begin{document}
%\title{Finding a Needle in Quantum Haystack with Deep Neural Networks}
\title{Artificial Intelligence for Quantum Matter: Finding a Needle in a Haystack}

\author{Khachatur Nazaryan}
%\email[Email: ]{khachnaz@mit.edu}
\thanks{These two authors contributed equally.}
\affiliation{Department of Physics, Massachusetts Institute of Technology, Cambridge, MA-02139,USA}

\author{Filippo Gaggioli}
\thanks{These two authors contributed equally}
\email[\\Email: ]{gfilippo@mit.edu}
\affiliation{Department of Physics, Massachusetts Institute of Technology, Cambridge, MA-02139,USA}

\author{Yi Teng}
\affiliation{Department of Physics, Massachusetts Institute of Technology, Cambridge, MA-02139,USA}

\author{Liang Fu}
\email[Email: ]{liangfu@mit.edu}
\affiliation{Department of Physics, Massachusetts Institute of Technology, Cambridge, MA-02139,USA}

\date{\today}

\begin{abstract}
Neural networks (NNs) have great potential in solving the ground state of various many-body problems. However, 
several key challenges remain to be overcome before NNs can tackle problems and system sizes inaccessible with more established tools. Here, we present a general and efficient method for learning the NN representation of an arbitrary many-body complex wave function from its $N$-particle probability density and probability current density and successfully test on (non-Abelian) fractional quantum Hall states and chiral BCS wavefunction.
Having reached overlaps as large as $99.9\%$, we employ our neural wave function for pre-training to effortlessly solve the fractional quantum Hall problem with Coulomb interactions and realistic Landau-level mixing for as many as $25$ particles and uncover distinctive features of the edge. 
Our work demonstrates efficient, scalable and accurate simulation of highly-entangled quantum matter using general-purpose deep NNs enhanced with physics-informed initialization.  
\end{abstract}
\maketitle

\textit{Introduction --} 
A fundamental challenge in many-body physics is the astronomical size of the Hilbert space: the number of complex amplitudes needed to completely specify a $N$-particle quantum wave function grows so quickly with $N$ that even modest systems outrun data storage and brute-force algorithms. Quantum computers could in principle solve certain quantum many-body problems efficiently, but with today’s noisy intermediate scale quantum processors, much of this promise is yet to be fulfilled. Recently, the artificial intelligence (AI) boom opened a different path \cite{Carleo_2017, Carrasquilla_2017, Glasser_2018, Carleo_2019, Luo_2019, Hermann_2020, Pfau_2020, Li_2022,Wilson_2023, Roth_2023,Cassella_2023,Pescia_2024, Lou_2024, Kim_2024,  Luo_2023, Smith_2024}: representing complex quantum wave functions with neural networks %which can capture essential quantum correlations through 
containing a tractable set of parameters and 
finding accurate approximation to ground states with present-day computing resources. 

Can a neural network architecture {\it accurately} and {\it efficiently} capture the vast variety of many‑body ground states of diverse quantum phases of matter (such as magnets, superconductors and topological materials)?  To grasp the scale of the challenge, recall that the complex wave function of a single particle in two spatial dimensions can be rendered as a colourful image whose intensity encodes amplitude $|\psi(\mathbf r)|$ and hue encodes phase $\varphi(\mathbf r)$. Learning the wave function of $N$ particles amounts to learning to generate a “hyper-image” that inhabits a $2N$‑dimensional configuration space. %—a task whose size grows exponentially with $N$ and which vividly illustrates the expressive burden placed on any finite neural architecture.
% In addition, approximation‑theory results—most notably those of Poggio and co‑workers \cite{Mhaskar_2016, Poggio_2024}—suggest that deep networks can succeed only when the target function is sufficiently “sparse” in a suitable compositional basis. 
%It is far from obvious that generic highly‑entangled quantum states meet this criterion, and their potential lack of sparsity, summed with the sheer size of the Hilbert space, could represent a fundamental obstruction to neural‑network quantum states (NQS).

As a concrete measure of the expressive power of neural networks, consider the \textit{needle‑in‑a‑haystack} problem: training a neural network to reproduce a target many‑body wave function $\ket{\psi_\text{ref}}$ that resides in the vastness of the Hilbert space. Success on this task would yield substantial rewards. It can be used for pre-training purpose to initialize networks at physically informed starting point, accelerating subsequent  ground state search by energy minimization in neural network variational Monte Carlo (NN-VMC).  
In addition, %a high-fidelity encoding of quantum states would allow NNs to address new fundamental questions, such as the information‑theoretic compressibility and entanglement structure of many-body wave-functions, 
training a neural network on a library of reference wavefunctions opens the door to data-driven  transfer-learning applications, such as predicting the electronic properties of a novel molecule from existing ones. Finally, low wavefunction fidelities currently limit the application of neural quantum states to studying the unitary dynamics of large quantum systems \cite{Hendry_2019, Gutierrez_2022, Sinibaldi_2023, Gravina_2025}. Developing a general method for achieving large overlaps would therefore advance important applications of AI for quantum dynamics \cite{Medvidovic_2021, Nys_2024, VandeWalle_2025, Parnes_2025}.  

While this needle‑in‑a‑haystack task %for neural network 
is easy to understand, it is by no means easy to achieve. % a neural network to reproduce a reference wave function $\psi_\text{ref}$ poses a formidable challenge. 
Even for a small system, almost all $N$-particle wave functions have vanishing overlap with the target  $\psi_\text{ref}$, and direct maximization of $|\!\braket{\psi_\text{ref}|\psi}|^2$ via gradient descent 
%succeeds
is extremely challenging. %The difficulty is compounded when $\ket{\psi_\text{ref}}$ is a highly-entangled state as in topological phases of matter, or when the complex wave function has an intricate phase pattern. 
To date, a general method for representing non-trivial target wave functions using neural networks is lacking.%, if not for simple non-interacting $\psiref$ \FG{(add citations)}.

Last but not the least, quantum statistics of identical particles imposes a fundamental constraint in their wave functions $\psi_\text{ref}({\bf r}_1, ..., {\bf r}_N)$, which must be anti-symmetric under the permutation of any two particles in Fermi systems. To comply with this condition, various Fermi neural network architectures have been introduced for electron systems in continuous space  
\cite{Hermann_2020, Pfau_2020, Li_2022,Wilson_2023, Cassella_2023,Pescia_2024, Lou_2024, Kim_2024,  Luo_2023, Smith_2024}. Compared with standard neural networks, their expressive power and training protocol are much less studied or benchmarked. The needle-in-a-haystack task would provide an objective ``score'' for the performance of Fermi neural network architectures.   

% \FG{Break into two paragraph. talk about time reversal and magnets.
% 1st paragraph about method: for the first time we are able to represent the moste general wave functions including complex ones. (probability denisty and probability current). Stress the method.
% then in second paragraph, as an application we apply to laughlin and moore read (in the future superconductivity)}

In this work, we develop a general and efficient method for learning the neural network representation of many-body wavefunctions. 
%to represent a target many-body wave function using deep neural networks. 
To circumvent the problem plaguing direct overlap maximization, we introduce a new training objective that targets the probability density and probability current of $\psiref$.  Our method is naturally suited to learning complex-valued wave functions, which appear ubiquitously in magnetic, chiral and spin-orbit-coupled quantum systems.

We test our method on archetypal many-body wave functions: the Laughlin state  and the Moore Read state in fractional quantum Hall systems, which represent topological quantum liquids hosting fractionally charged quasiparticles (``anyons'') with Abelian and non-Abelian statistics, respectively, and the chiral $p$-wave BCS wave function for spin-polarized electrons, which hosts exotic Majorana fermions at its edges. A {\it general-purpose} Fermi neural network architecture based on self-attention is employed for all these tasks, without prior knowledge of quantum Hall physics or superconductivity. Remarkably, our {\it unsupervised} learning method successfully finds neural network representations of these highly-entangled wavefunctions, reaching overlaps as large as $99.9\%$ for as many as $25$ particles. %\FG{using different architecture sizes}. %\FG{We systematize our analysis by comparing the overlaps for different particle numbers and architecture sizes.}

Using these trained neural networks and performing NN-VMC \cite{Carleo_2019_review, Hermann_2023_review} for energy minimization, we effortlessly solve the ground state of the fractional quantum Hall (FQH) system for $N=25$ particles with Coulomb interaction and realistic Landau-level mixing. We find that the FQH droplet displays long-ranged density oscillations that extend away from the disk boundary. %, suggesting that $1$D descriptions of the chiral edge modes are insufficient.} 
This success demonstrates the power of our method for pretraining on physically motivated ansatz, enabling fast and accurate neural network solution of strongly correlated electron systems.  

% By solving the “needle” problem, we (i) validate the expressive power of deep neural network for representing topologically ordered, highly entangled states; (ii) provide a robust and efficient pre‑training scheme based on physically motivated initialisation, which enables and accelerates neural network solution of strongly correlated electron systems, such as fractional quantum Hall states and chiral superconductors.
% %In a broader context, our method opens the door to data‑driven transfer‑learning pipelines that extrapolate to unseen quantum systems.

%

\textit{Loss functions --} 
% In modern neural-networks, the loss function $L$ provides the quantitative objective that shapes how the model learns. Gradient-based optimisation then adjusts network parameters to minimise this scalar criterion. 
% Given a well-defined objective, different loss functions can guide the same learning process.
%However, any
% Every successful loss function must satisfy two important conditions: (i) a reduction of $L$ must coincide with progress towards the underlying objective, and (ii) the optimisation landscape must deliver consistently informative gradients—avoiding flat regions or “barren plateaus”—so that the optimiser can reliably descend toward a high-quality local minimum of $L$.
For the needle problem, the key figure of merit is the fidelity (or squared overlap)
$F = \left|\braket{\psiref|\psitheta}/\|\psiref\|\|\psitheta\|\,\right|^2$, with the wave function norm defined as $\|\psi \|^2=\braket{\psi|\psi}$.
The fidelity naturally provides us with a simple choice for the loss function $\LF = 1 -F$. In the form of a Monte Carlo expectation value (see the supplementary material (SM) \cite{supplementary} for details), this reads
\begin{equation}\label{eq:lossF}
\LF = 1-\frac{\left|\int \mathrm{d}\Rb \,|\psitheta(\Rb)|^2 \psiref(\Rb)/\psitheta(\Rb)\right|^2/\mathcal{N}^2}{\int \rm{d}\Rb\,|\psitheta(\Rb)|^2 \,|\psiref(\Rb)/\psitheta(\Rb)|^2/\mathcal{N}},
\end{equation}
where $\mathcal{N} = \int \rm{d}\Rb\,|\psitheta(\Rb)|^2$ 
and the integration variable $\Rb=(\rb_1,\cdots,\rb_N)$ spans the $\mathbb{R}^{2N}$ coordinate space of $N$ particles in $2$D.
However, the overlap of $\psiref$ with another wave function is in general exponentially small, implying that the gradients of $\LF$ will be unable to guide the neural network across the optimization landscape for all but the smallest system size.
In the case of real $\psiref$, the exponentially small gradients can be “amplified”  by working with the logarithms of the wave functions \cite{Matthew_2019}. %, as typically done in quantum chemistry for the pre-training of neural networks \cite{to be added}. 
This simple fix, however, is not sufficient in the case of truly complex $\psiref$. %, such as those characteristic in magnetic, chiral and spin-orbit-coupled materials. 

% As anticipated in the introduction, however, 
% the overlap of $\psiref$ with a random wave function in the Hilbert space is in general exponentially small, and even more so in the presence of complex phase patterns. The fidelity loss in Eq.\ \eqref{eq:lossF}, and all the variations constructed by directly minimization of some function of the overlap $|\braket{\psiref|\psitheta}|$,
% will therefore be unable to initiate the gradient descent in all but the smallest Hilbert spaces.

While the modulus of the wave function represents the $N$-particle probability density $\rho(\Rb) = |\psi(\Rb)|^2/\mathcal{N}$ and is closely related to physical observables, the phase $\varphi$ is a more subtle quantity that cannot be directly accessed experimentally and is only defined up to a constant.
The  phase gradient $\nabla \varphi$, on the other hand, encodes important information about the current flowing within the system: 
$\bf{j} \propto  \rho \nabla \varphi$ represents the probability current density.  %carried by the $\ell$-th particle in configuration $\Rb$. 
Motivated by this observation, we introduce a new loss function that consists of two parts, $\Loss_\rho$ and $\Loss_j$, respectively designed to minimize the difference in the particle density and the phase gradients between the trial and target wave functions.

The density loss function $\Loss_\rho$ is inspired by the Kullback–Leibler divergence \cite{Kullback_1951} that measures the distance between the probability distributions $|\psitheta|^2$ and $|\psiref|^2$, and reads
\begin{equation}\label{eq:lossrho}
    \Loss_\rho = \frac{1}{\mathcal{N}}\int \rm{d}\Rb\,|\psitheta(\Rb)|^2 \left(\ln|\psitheta(\Rb)/\psiref(\Rb)|^2 \right)^2.
    % \Loss_\rho = \frac{\int \rm{d}\Rb\,|\psitheta(\Rb)|^2 \left(\ln|\psitheta(\Rb)|^2 -  \ln|\psiref(\Rb)|^2 \right)^2}{\int \rm{d}\Rb\,|\psitheta(\Rb)|^2}.
\end{equation}
As discussed above, this particular choice of $\Loss_\rho$ has the advantage that it retains sensitivity when either of $|\psitheta|^2$ and $|\psiref|^2$ is very small, thanks to the difference between logarithms.
The current loss function $\Loss_j$ instead takes the simple form 
\begin{equation}\label{eq:lossj}
    \Loss_j =  \frac{1}{\mathcal{N}}\int \rm{d}\Rb\,|\psitheta(\Rb)|^2%|\psiref|^2 
    \sum_\ell\left|\nabla_\ell\, \phitheta(\Rb) -\nabla_\ell\, \phiref(\Rb) \right|^2, 
    % \Loss_j =  \frac{\int \rm{d}\Rb\,|\psitheta(\Rb)|^2%|\psiref|^2 
    % \sum_l\left|\nabla_l \phitheta(\Rb) - \nabla_l \phiref(\Rb) \right|^2}{\int \rm{d}\Rb\,|\psitheta(\Rb)|^2}, 
\end{equation}
where $\nabla_\ell$ is the gradient with respect to the $\ell$-th particle position $\rb_\ell$, while $\phitheta$ and $\phiref$ are the phases of $\psitheta$ and $\psiref$ \footnote{Equation \eqref{eq:lossj} is written for electron systems in continuous space, but an analogous expression can be obtained for systems on a lattice. 
}. 
Due to the presence of the gradient, the current loss function \eqref{eq:lossj} captures the spatial variation of the phase 
(which is physically observable) and prevents the fragmentation of $\phitheta$ into local patches that differ by integer multiples of $2\pi$.
Moreover, the non-local character of the spatial derivatives allows the loss function to probe the low-density regions otherwise inaccessible to the Monte Carlo sampling. These properties make $\Loss_j$ very well-suited for capturing the phase pattern of wave functions that display singularities such as vortices, as we will show below.
%-- the benefits of this will become evident below.

The total loss function is finally obtained by summing Eqs.\ \eqref{eq:lossrho}-\eqref{eq:lossj},
\begin{equation}\label{eq:totalLoss}
\Loss =\Loss_\rho + \alpha \Loss_j.    
\end{equation}
The coefficient $\alpha>0$ is an important hyperparameter that balances the relative weight of the density- and current loss functions, and needs to be optimized depending on the choice (and normalization) of $\psiref$.
%Once a good overlap is reached by training the network with this loss function, the quality of the neural wave function can further be improved by continuing the training with the fidelity loss \eqref{eq:lossF}, as discussed  ... .

Our method is applicable to both Bose and Fermi systems. In the rest of this work, we will demonstrate its effectiveness for Fermi systems. %, where the many-body wave function must be antisymmetric under the exchange of two particles.    

\begin{figure}
    \includegraphics[width=0.9\linewidth]{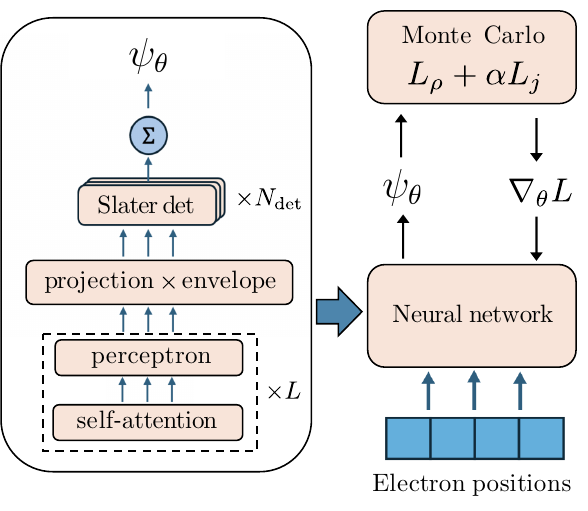}
    \caption{\textbf{Fermionic neural network and VMC:} Illustration of our fermionic attention-based architecture (left), and its role inside the NN variational Monte Carlo (right). }
    \label{fig:architecture}
\end{figure}

\begin{figure*}
    \includegraphics[width=0.99\linewidth]
    {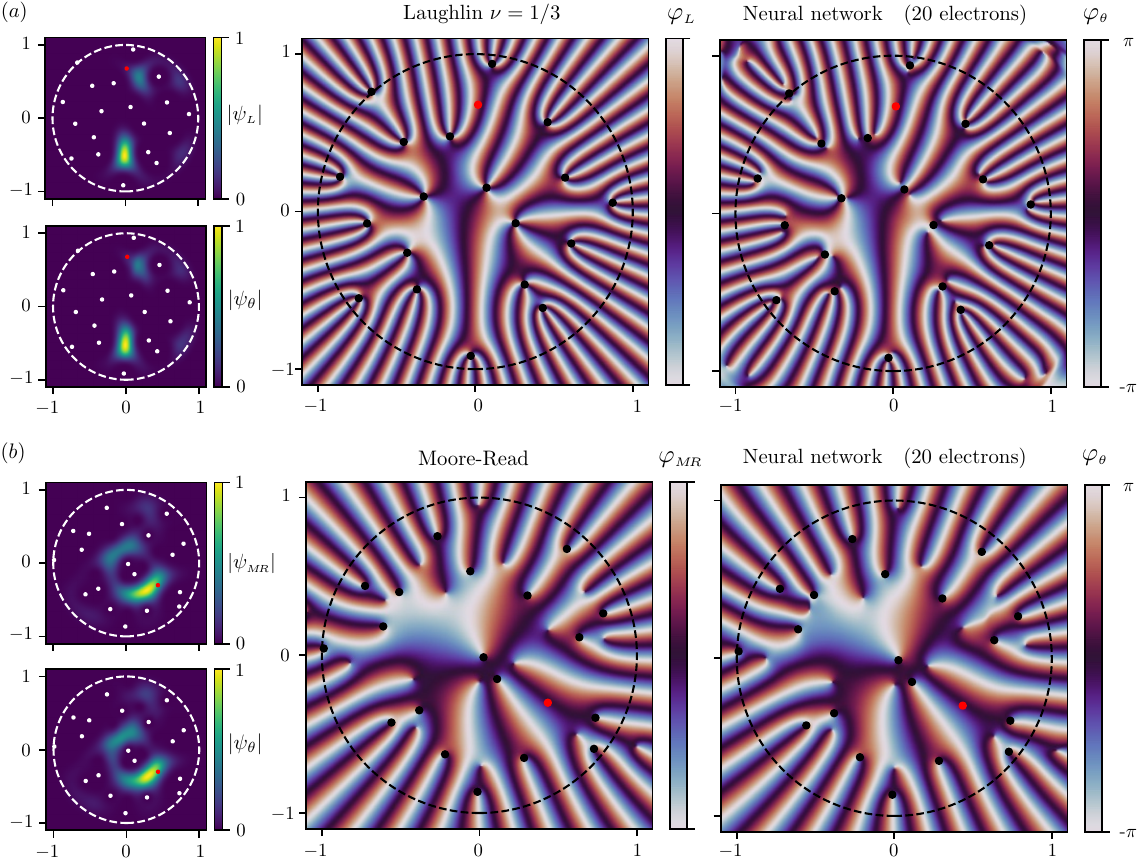}
    \caption{\textbf{Laughlin and MR wave-functions:} Comparison of the wave functions for the Laughlin $(a)$ and MR $(b)$ state, Eqs.\ \eqref{eq:psiL}-\eqref{eq:psiMR},  with the output of the neural network ($2$ self-attention layers and $4$ determinants). %($L=2$ self-attention layers and $N_\text{det}=4$ determinants). 
    These plots are obtained by keeping $N - 1$ particles at fixed positions (black/white dots), obtained from Monte Carlo sampling, and moving the remaining particle away from its "original" position (red dot) across the $2$D plane (positions in units of the droplet radius $R_{\scriptscriptstyle L} = \sqrt{6N\ell_M^2}$ $(a)$ and $R_{\scriptscriptstyle M\!R} = \sqrt{4N\ell_M^2}$ $(b)$, with $\ell_M^2 = \phi_0/2\pi H$ the magnetic length associated to the out of plane field $H$). 
    }
    \label{fig:laughlin_MR}
\end{figure*}

\textit{Fermionic neural network --}
%We will now illustrate the effectiveness of our novel method using a general fermionic neural network.
A number of neural network architectures have been developed to represent fermion wave functions in continuous space.  %\cite{Pfau_2020, Hermann_2020, Li_2022, Cassella_2023, Wilson_2023, Pescia_2024,  Lou_2024, Kim_2024,  Luo_2023, Smith_2024}. 
Commonly used architectures,         
%The workflow of fermionic NNs, which is shared across different architectures 
such as FermiNet \cite{Pfau_2020} and PauliNet \cite{Hermann_2020} and self-attention based neural networks \cite{vonGlehn_2023,Geier_2025}, 
take the particle coordinates as input, combine them into a set of ``orbitals'' that depend on the positions of all electrons, and finally assemble these {\it many-electron orbitals} into Slater determinants to construct an anti-symmetric wave-function that respects Fermi statistics.  
By incorporating multiparticle correlations into many-electron orbitals, these neural ansatz go beyond Hartree-Fock approximation
and can capture the ground states of various correlated electron systems, %. They provide powerful variational ansatz for the ground state of 
as demonstrated for atoms, molecules and solids \cite{Hermann_2020,Pfau_2020,vonGlehn_2023, Li_2022, Geier_2025}.  %and 
%In condensed matter physics, they have recently been used to study various many-body problems including moiré materials \cite{Geier_2025, Di_2025, Li_2025}. %, attaining accurate results.

%\addKN{Our attention-based neural network  $->$ We use an attention based neural network} 

Our neural network ansatz is inspired by the transformer architecture originally proposed in the context of large language models \cite{Vaswani_2017}, and uses self-attention mechanism to capture electron correlations \cite{vonGlehn_2023,Geier_2025}. As illustrated in Fig.\ \ref{fig:architecture},  
it consists of a stack of self-attention and perceptron layers, repeated $L$ times, that takes  the electron positions $\rb_j$ as input and outputs vectors that, after projection and convolution with a simple Gaussian envelope, create the generalized single-particle orbitals $\phi^{(k)}_i\left(\rb_j,\lbrace{\rb_{/j}\rbrace}\right)$. % with $k \leq N_\text{det}$. 
These are finally combined into $N_\text{det}$ Slater determinants, 
whose sum constitutes the antisymmetric fermionic neural wave function 
\begin{equation}
    \psitheta(\rb_1,\cdots, \rb_N) = \sum_k^{N_\text{det}} \det\left[\phi_i^{(k)}\left(\rb_j,\lbrace{\rb_{/j}\rbrace}\right)\right],
\end{equation}
that is parametrized by the network weights $\theta$.

Having constructed the wave function, $\psitheta$ is used as a variational ansatz in the VMC algorithm, where the desired loss function $L$ is evaluated by means of Monte Carlo techniques.
The gradients $\nabla_\theta L$ of the loss function are finally passed back to the neural network to update the weights $\theta$ via standard backpropagation after each training step. %\addKN{This sentence does not read well.} 

%Due to the absence of artful but restrictive artifacts, such as tailor-made envelope functions or Jastrow factors, our neural architecture is general and unbiased, holding the promise to be an effective fermionic wave function approximator across many different phases of matter.

\textit{Results: fractional quantum Hall -- }
The fractional quantum Hall effect is an archetypal problem of many-body condensed-matter physics, and showcases an intricate interplay between strong electronic correlations and non-trivial topology. Much of the field’s progress has come from remarkably insightful trial wave-functions, most famously Laughlin’s \cite{Laughlin_1983}
\begin{equation}\label{eq:psiL}
    \psiL = \prod_{i<j}(z_i - z_j)^3 \exp(-|z_i|^2/4),
\end{equation}
which captures the essential physics of the true ground state at filling $1/3$. 
The Laughlin state supports charge-$1/3$ quasiparticles that are Abelian anyons. 
Another celebrated trial wavefunction is the Moore–Read Pfaffian state \cite{Moore_1991}
\begin{equation}\label{eq:psiMR}
    \psiMR = \text{Pf}\left[\frac{1}{z_i - z_j}\right]\prod_{i<j}(z_i - z_j)^2 \exp(-|z_i|^2/4),
\end{equation}
which supports charge-$1/4$ quasiparticles that have non-Abelian statistics.   
%While both $\psiL$ and $\psiMR$ were first discovered in the context of the FQH,
The Moore-Read wave function \eqref{eq:psiMR} can be viewed as a paired state of composite fermions, 
and hence belongs to a different and more exotic ``universality class'' than the Laughlin state.

The wave functions $\psiL$ and $\psiMR$ are shown in Fig.\ \ref{fig:laughlin_MR}$\,(a)$ and $(b)$ as a function of the position of a single particle, while the remaining $N-1$ are fixed (white/black dots) in a typical configuration that was sampled from  Eqs.\ \eqref{eq:psiL}-\eqref{eq:psiMR} using Monte Carlo methods ($N=20$ for both Laughlin and Moore-Read).
The absolute values $|\psiL|$ and $|\psiMR|$ (top left panels) have the spatial profile characteristic of strongly correlated systems: the position of the ``last''  particle is strongly constrained by every other particle's coordinates. 
The phases $\phiL$ and $\phiMR$ (central panels), on the other hand, display an intricate pattern: 
the Laughlin state generally features vortices with $6\pi$ phase winding where two particles coincide, 
while the phase pattern of the Moore-Read state is even more subtle.   
%that results from the combination of time-reversal symmetry breaking with strong electron-electron interactions.
The highly complex nature of these model wavefunctions, which embodies the universal physics of 
the fractional quantum Hall effect, makes them the ideal ``needles'' for testing our neural network learning method.

Evaluating $|\psitheta|$ and $\phitheta$ on the same pair of electron configurations using our attention-based neural network, we obtained the results shown in the remaining panels of Fig.\ \ref{fig:laughlin_MR} $(a)$ and $(b)$. 
For these plots, we trained our NN using the loss function \eqref{eq:totalLoss} (see the SM \cite{supplementary} for details on the network and the training protocol). 
% As shown in Fig.\ \ref{fig:laughlin_MR}, the resulting wave function $\psitheta$ precisely captures the strongly correlated features of the electron density distribution, which is highly localized in a corner of the Laughlin droplet.
% Even more strikingly, the neural network prediction for $\phitheta$ almost perfectly reproduces the intricate phase patterns $\phiL$ and $\phiMR$ not just in correspondence of the high density regions, which dominate the Monte Carlo averages, but also far away and close to the nodes of the target wave-functions, where $|\psiL|^2$ and $|\psiMR|^2$ are vanishingly small.
% We believe that this is a direct consequence of the presence of the spatial derivatives in the current loss function $\eqref{eq:lossj}$, whose action endows the training process of nonlocal information about the structure of the wave function away from the hig-density regions that are otherwise statistically inaccessible to direct Monte Carlo sampling.
The modulus of $\psitheta$ faithfully captures the strongly correlated electron density. %, which is sharply confined to a corner of the Laughlin droplet. 
Even more remarkable is the network’s phase prediction, $\phitheta$, which accurately reproduces the intricate patterns of $\phiL$ and $\phiMR$ not only in the high-density regions that dominate the Monte-Carlo averages, but also in the low-density areas near the nodes of the target wave functions, where $|\psiL|^{2}$ and $|\psiMR|^{2}$ are vanishingly small. The accuracy of $\phitheta$ close to these points is a beneficial consequence of the non-locality of $\Loss_j$, as anticipated in the discussion below Eq.\ \eqref{eq:lossj}.

\begin{figure}
    \includegraphics[width=0.99\linewidth]
    {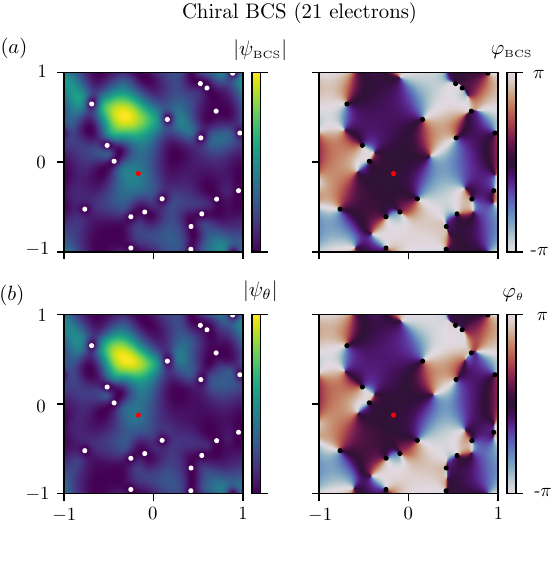}
    \caption{\textbf{Chiral BCS wave-function:} Comparison of the reference wave function \eqref{eq:chiralBCS_odd} (odd particle number) with the output of the neural network ($5$ self-attention layers and $4$ determinants). 
    %As for Fig.\ \ref{fig:laughlin_MR}, these plots are obtained by keeping $N - 1$ particles at fixed positions and moving the remaining particle across the $2$D square toroidal supercell. 
    For evaluating $\psiBCS$, we 
    set the chemical potential $\mu = 0.5 \times (k_F^2/2m)$ (Fermi momentum $k_F = \sqrt{4\pi N/ L^2}$) and the gap function to $\Delta_k = \mu (k/k_F)\, \exp\!\left[-(k/2 k_F)^2\right]$, using a soft UV cutoff to regularize the Fourier summation.
    }
    \label{fig:BCS_needle}
\end{figure}

\textit{Results: chiral superconductivity -- }The chiral $p$-wave superconducting wave function is constructed by projecting the BCS paired state of spinless electrons with periodic boundary conditions onto a fixed number of particle sector. For an even number of particles on a square torus of size $L$, this reads \cite{Read_2000}
\begin{equation}\label{eq:chiralBCS_even}
\psiBCS = \text{Pf}\left[g(z_i - z_j)\right], \quad g(z) = L^{-2}\sum_{k}e^{i\bar{k}z}\,g_k,
\end{equation}
with the pair-correlation function $ g(z)$ being the inverse Fourier transform of $g_k = (\xi_k - E_k)/\bar{\Delta}_k$ and $k = k_x + i k_y$ the reciprocal lattice vectors.
In the above, we introduced the free electron and quasiparticle energies $\xi_k = k^2/2m - \mu$ and $E_k = \sqrt{\xi_k^2 + |\Delta_k|^2}$, along with the gap function for complex $p$-wave pairing $\Delta_k \propto  k$.
From these definitions, one sees that the small-$k$ behavior of depends crucially on the sign of the chemical potential $\mu$: for $k\to 0$, $g_k$ diverges for $\mu > 0$ and vanishes for $\mu <0$, with a (topological) transition at $\mu = 0$ \cite{Read_2000}. The positive and negative cases correspond to weak- and strong-pairing phases because the spatial profile of the Cooper pairs (as quantified by the long-distance behavior of $g(z)$) changes from power-law decay to exponential decay, respectively.

The divergence of $g_{k\to 0}$ in the weak-pairing phase implies that, for a system with periodic boundary conditions, the chiral superconductor favors an odd number of particles with a single, unpaired, electron occupying the $k = 0$ mode.
In this case, the BCS wave function consists of the superposition of all possible electron pairs given by Eq.\ \eqref{eq:chiralBCS_even} multiplied by the constant $k = 0$ free electron orbital,  anti-symmetrized over the unpaired particle index,
\begin{equation}\label{eq:chiralBCS_odd}
\psiBCS = \sum_{l=1}^N(-1)^l\,\text{Pf}\left[g(z_{i\neq l} - z_{j\neq l})\right]/L,
\end{equation}
with the pair correlation function $g(z)$ now constructed exclusively from the $g_k$ components with momentum $k\neq0$, see Eq.\ \eqref{eq:chiralBCS_even} above.

The rich topological properties of the weak-pairing phase, which gives rise to Majorana modes at its boundaries,  make the chiral BCS state \eqref{eq:chiralBCS_odd} a compelling and ambitious target for applying our needle method.
The results are shown in Fig.\ \ref{fig:BCS_needle}: as for the FQH wave functions \eqref{eq:psiL}-\eqref{eq:psiMR}, the neural network reaches very high overlaps $> 99\%$ for as many as $21$ particles and succeeds in capturing the convoluted phase pattern $\phiBCS$ characteristic of the chiral BCS state \eqref{eq:chiralBCS_odd}.
At the same time, the neural network reproduces closely the strongly delocalized amplitude profile $|\psiBCS|$, confirming the strength and versatility of the loss function \eqref{eq:totalLoss} across a variety of different many-body wave functions.

 \begin{figure}
    \includegraphics[width=\linewidth]{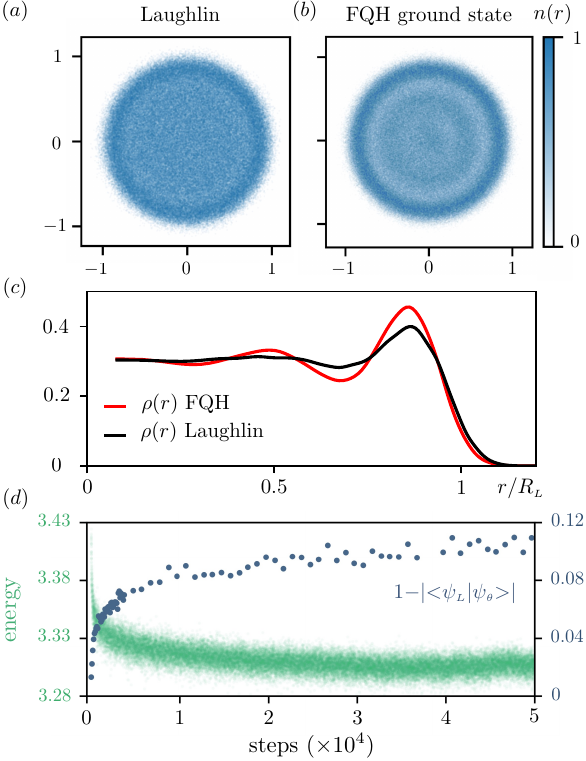}
    \caption{\textbf{FQH ground state:} 
    Spatial density profiles of the Laughlin droplet $(a)$ and FQH ground state for mixing parameter $\lambda = 1$ $(b)$ ($N=25$ and positions expressed in units of $R_{\scriptscriptstyle L}$). $(c)$ In the FQH ground state, Coulomb forces induce long-ranged oscillations of the density $\rho(r)$ that are slowly decaying on the scale of the system size, contrary to the exponential decay for the Laughlin state. $(d)$ Evolution of the variational energy (green) and ``distance'' from the Laughlin state (blue), as measured by $1 - |\braket{\psiL|\psitheta}|$. As the energy gradually decreases, the wave function $\psitheta$ diverges away from $\psiL$.} %(the dashed line provides a guide for the eye).}
    \label{fig:FQH}
\end{figure}

\textit{Application to pre-training --}
Recently, self-attention-based neural networks have demonstrated impressive success in finding the ground state of the fractional quantum Hall problem with Landau-level (LL) mixing for system sizes up to twelve particles, outperforming 
traditional approaches, such as exact diagonalization (ED) with Landau level truncation \cite{Teng_2025, Qian_2025}. 
Indeed, while ED is fundamentally limited by the exponential growth of the Hilbert space, %making large particle numbers intractable, 
 neural network based variational method can in principle avoid this bottleneck and attain accurate solution for large systems. 
 However, as the system size increases, the optimization landscape becomes increasingly complex and the neural network training can easily fail to converge, %effectively getting lost despite 
 even with substantial computational time and resources.

By pre-training our neural network to maximize the overlap with $\psiL$, we are now able to overcome this problem and efficiently solve the FQH problem with strong LL mixing for an unprecedented system size. For $25$ electrons (which is inaccessible to even ED within the lowest Landau level), 
the corresponding results for mixing parameter $\lambda = e^2/4\pi \varepsilon_0\epsilon\ell_B = 1$  (see SM \cite{supplementary} for details) are shown in Fig.\ \ref{fig:FQH}, where we compare the spatial density profile for the Laughlin $(a)$ and FQH $(b)$ droplet in disk geometry. By studying their radial profiles $(c)$, it becomes evident that the long-ranged Coulomb repulsion induces slowly-decaying oscillation in the charge density away from the edge of the disk,
consistent with previous studies on much smaller system sizes \cite{Tsiper_2001, Wan_2003, Hu_2019,Teng_2025}.
This indicates that the edge of the FQH droplet has an extended effect on the bulk, suggesting that the idealized chiral Luttinger liquid description of the edge state \cite{Wen_1990} may not be applicable to FQH systems with Coulomb interaction \cite{Jolad_2009}.

The decrease in energy, of the order of few percents when compared to the initial Laughlin state, is shown in panel $(d)$ (small green dots), along with the rapid evolution of $\psitheta$ away from $\psiL$ as measured by the ``distance'' $1 - |\braket{\psiL|\psitheta}|$ (blue dots), which goes from $\approx 1\%$ to $\approx 12\%$. 
These results clearly show the important distinction between the Laughlin wavefunction and the actual Coulomb ground state.  
On the other hand, 
during the entire training process, the total angular momentum of the system remained very close ($\approx 901.80$) to the integer value of $900$ for the Laughlin state with $25$ particles. %, %demonstrating the robust quantization of topological quantum number in the fractional quantum Hall liquid.  
In the SM \cite{supplementary}, we analyze the scaling of the energy per particle with the system size.

Altogether, these results demonstrate that our self-attention NN is capable of solving the FQH problem with realistic LL mixing for large system sizes with modest computational resources \cite{supplementary}, once the 
neural network is appropriately pre-trained. 
For comparison, while Laughlin and Moore-Read model wavefunctions are faithfully described by matrix product states (MPS)  \cite{Zaletel_2012, Estienne_2013}, such MPS representations do not extend to the ground state of realistic FQH systems with Landau level mixing. As a result, we believe that the success of our NN method truly stands out.

 \begin{figure}
    \includegraphics[width=\linewidth]{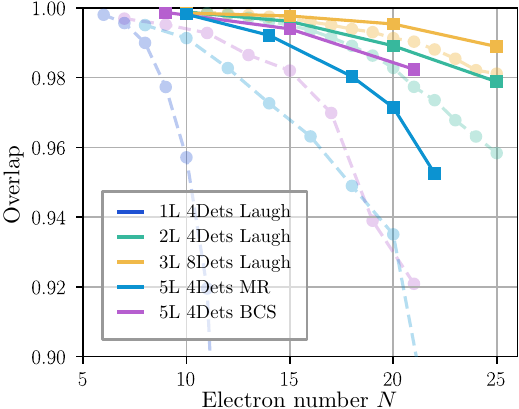} 
    \caption{\textbf{Scaling analysis:} 
    Overlap with the Laughlin wave function as a function of the particle number, for three different architectures (dark blue, green and yellow). The light blue curve is for the Moore-Read state, and the purple for the chiral BCS state. The two different type of lines represent two training protocols: minimizing only loss function \eqref{eq:totalLoss} (dashed); further improving the result by optimizing fidelity loss \eqref{eq:lossF} (full).}
    \label{fig:scaling}
\end{figure}

\textit{Training protocol and scaling analysis --}
To conclude our discussion, we go back to the needle problem and discuss the scaling of the overlap as a function of particle number for three different self-attention architectures, with varying number of layers ($L=1,2,3,5)$ %\addKN{(dark blue, green, yellow and purple/blue)} 
and Slater determinants ($N_\text{det} =4, 8$).
At the same time, we compare two different training protocols: a shorter one (faint dashed lines, circles), where the overlap is learned entirely by minimizing the loss function \eqref{eq:totalLoss} for a fixed number of steps with the hyper-parameter $\alpha$ gradually increasing from zero to unity; and a longer one (solid lines, squares), where in a second part of the training the fidelity loss \eqref{eq:lossF} is directly minimized. Further details are provided in the SM \cite{supplementary}.

A key ingredient of our training protocol is a \textit{transfer-learning method} that exploits the size-independent nature of the self-attention layers. Training is initiated at a small system size, where the network is optimized to reproduce the target wave function, and the resulting weights are then reused to initialize the 
$N+1$-particle problem (cf SM \cite{supplementary}). Because the self-attention layers are independent of the particle number, their parameters can be transferred directly, while only a minimal subset of size-dependent parameters—associated with parts of the orbitals and the envelope—must be reinitialized.
By iterating sequentially to larger systems, this transfer learning approach preserves the high-quality wave functions learned at smaller $N$ and dramatically reduces the optimization cost, providing an efficient and robust route to large system sizes that would be prohibitive to tackle otherwise.

We now discuss the scaling of the different wave function overlaps as a function of the particle number $N$, for different architectures. As shown in Fig.\ \ref{fig:scaling}, the $2$- and $3$-layer architectures (green and yellow, number of parameters $\approx0.8$ and $1.3\times 10^6$ respectively) excel at reproducing the Laughlin wave function for up to $25$ particles, the largest system size studied in this work. 
At the same time, the Moore-Read state for $22$ particles can be faithfully reproduced with $\approx 95\%$ overlap using the $5$-layers architecture (blue, number of parameters $\approx 0.8\times10^6$). Finally, for chiral BCS state with 21 particles, we reach over $98\%$ overlap (purple).
This favorable scaling highlights the expressive power of self-attention networks for capturing quantum phases of matter 
and suggests that our method for deep learning a target many-body wavefunction  
is well-suited to tackle even larger system sizes.

\textit{Discussion --}
The versatility, accuracy and efficiency of neural networks are the crucial ingredients underpinning the rapid development 
of AI-based methods across different branches of condensed matter physics.
Our work expands the AI-for-quantum horizon by introducing a general unsupervised learning method to represent arbitrary wave functions, demonstrating the expressive power of self-attention neural networks. 
By targeting the Laughlin and Moore-Read wave functions, which describe archetypal topologically ordered many-body states, we demonstrate high overlaps $> 99\%$ for as many as 25 particles using a simple self-attention NN without prior knowledge of quantum Hall physics. 
Performing NN-VMC for energy minimization with these pre-trained neural networks, we effortlessly solve the ground state of the fractional quantum Hall system with Coulomb interaction and strong Landau-level mixing for unprecedented system sizes.

Our general method provides a useful tool for pre-training wave functions, opening the door to many applications of neural networks to quantum condensed matter physics, 
in particular many-body systems in continuous space
where traditional methods suffer from band-projection or discretization error.
Of particular interest are the investigation of composite fermion states \cite{Jain_2007_book, Pu_2018, Gattu_2025} and the study of non-Abelian fractional quantum Hall states, moir\'e fractional Chern insulators, and 
chiral superconductivity \cite{Han_2024, Xu_2025, Li_2025}. More broadly, our results demonstrate that fast, accurate simulation of complex quantum matter can be achieved by enhancing deep NNs with physics-informed initialization, while retaining their expressivity and accuracy. 
%, and move us an inch closer to AI supremacy in condensed matter physics.

\textit{Acknowledgments --} This work was primarily supported by National Science Foundation (NSF) Convergence Accelerator Award No. 2235945. We acknowledge the MIT SuperCloud and Lincoln Laboratory Supercomputing Center for providing computing resources that have contributed to the research results reported within this paper. 
F.G. is grateful for the financial support from the Swiss National Science Foundation (Postdoc.Mobility Grant No. 222230).  
K.N. acknowledges the support from the NSF through Award No. PHY-2425180.  L.F. was supported by a Simons Investigator Award from the Simons Foundation.

\bibliography{biblio.bib}

\onecolumngrid
\newpage
\makeatletter 

\begin{center}
%\textbf{\large Supplementary materials for: ``{Artificial Intelligence for Quantum Matter:\\Finding a Needle in a Haystack} ''}
\textbf{\large Supplementary materials for:\\ ``Artificial Intelligence for Quantum Matter: Finding a Needle in a Haystack''}
\\[10pt]
Khachatur Nazaryan$^{1}$, Filippo Gaggioli$^{1}$, Yi Teng$^{1}$ and Liang Fu$^{1}$ \\
\textit{$^1$Department of Physics, Massachusetts Institute of Technology, Cambridge, MA, USA}
\end{center}
\vspace{20pt}

\setcounter{figure}{0}
\setcounter{section}{0}
\setcounter{equation}{0}

\renewcommand{\thefigure}{S\@arabic\c@figure}
\makeatother

 %\appendix 

These supplementary materials contain details about the loss functions, the training protocol and the FQH energy minimization problem presented in the main text.

\section{Non-linear loss function and gradient}

In conventional Variational Monte-Carlo (VMC) one minimizes the
expectation value of the Hamiltonian,
\[
\mathcal{H}[\psi_\theta]= \frac{
\langle\psi_\theta|\hat H|\psi_\theta\rangle}{\langle\psi_\theta|\psi_\theta\rangle} = E_{p_\theta}[E_L ], \quad E_L = \psi_\theta^{-1} (\hat{H} \psi_\theta),
\]
Throughout the manuscript, $\boldsymbol{R}$ denotes the coordinates of all the particles (configuration), and the expectation value
\(
\langle f\rangle_{p_\theta}\equiv
\int d\mathbf R\,p_\theta(\mathbf R)\,f(\mathbf R)
\)
is abbreviated as $E_{p_\theta}[f]$.

Thanks to the variational principle—this guarantees an upper bound to
the ground-state energy.  Crucially, the Hamiltonian \(\hat H\) acts
linearly on the trial wave-function \(\psi_\theta\), so both the loss
and its gradient inherit a particularly simple structure.
\[
\partial_{\theta}\mathcal{H}
  \;=\;
  E_{p_\theta} \bigg[ \,
      E_{L}\,\partial_{\theta}\log\psi^*_\theta
      \;+\;
      E_{L}^{*}\,\partial_{\theta}\log\psi_\theta
      \;-\;
      2\, E_{p_\theta}[ E_{L} ]\,
      \partial_{\theta}\log|\psi_\theta| \bigg].
\]
This approach allows for efficient and stable optimization.

In the present work we must go beyond linear operators and instead minimize
losses that are \textit{non-linear} functionals of \(\psi_\theta\).  
We introduced these loss functions in the main text, and include density-based loss and probability current-based loss.
These objectives do not factorize into a single application of \(\hat H\) and
therefore require a more delicate treatment of the
stochastic expectations, and gradient estimates.  The remainder of this
section derives explicit, Monte-Carlo-friendly expressions for both the losses
and their parameter gradients, providing the foundations for stable training
with non-linear objectives.

\subsection{Density loss}

\subsubsection{Loss calculation}

To retain informative signals deep in the low-overlap regime we use a
\textit{density–based} comparison and work with
probability densities
\(
p_\theta(\mathbf R)=|\psi_\theta(\mathbf R)|^2/N_\theta^2
\quad\text{and}\quad
p_{\mathrm{Ref}}(\mathbf R)=|\psi_{\mathrm{Ref}}(\mathbf R)|^2/N_{\mathrm{Ref}}^2,
\)
where the denominators normalize each distribution.

The Kullback–Leibler (KL) divergence is a natural candidate because it measures
the directed distance between densities and is strictly positive except at
perfect agreement.

\begin{equation}
\label{eq:kl_def}
\mathcal L_{\mathrm{KL}}
  =\frac{\int d\boldsymbol{R}\:|\psi_{\theta}(\boldsymbol{R})|^{2}\ln\left(\frac{|\psi_{\theta}(\boldsymbol{R})|^{2}/N_{\theta}^{2}}{|\psi_{\text{Ref}} (\boldsymbol{R})|^{2}/N_{\text{Ref}}^{2}}\right)}{\int d\boldsymbol{R}\:|\psi_{\theta}(\boldsymbol{R})|^{2}}
\end{equation}

For the normalization factors we can write,
\begin{align}
    &\frac{N_{\theta}^{2}}{N_{\text{Ref}}^{2}}=\frac{\int d\boldsymbol{R}\:|\psi_{\theta}(\boldsymbol{R})|^{2}}{\int d\boldsymbol{R}\psi_{\text{Ref}}^{*}(\boldsymbol{R})\psi_{\text{Ref}}(\boldsymbol{R})}=\frac{\int d\boldsymbol{R}\:|\psi_{\theta}(\boldsymbol{R})|^{2}}{\int d\boldsymbol{R}|\psi_{\theta}(\boldsymbol{R})|^{2}\frac{|\psi_{\text{Ref}}(\boldsymbol{R})|^{2}}{|\psi_{\theta}(\boldsymbol{R})|^{2}}}=\frac{1}{E_{p_\theta}\left[|\frac{\psi_{\text{Ref}}(\boldsymbol{R})}{\psi_{\theta}(\boldsymbol{R})}|^{2}\right]}
\end{align}

Eq.~\eqref{eq:kl_def} can be rewritten in this notation as
\begin{equation}
\label{eq:kl_split}
\mathcal L_{\mathrm{KL}}
 = E_{p_\theta}\!\Bigl[
     \ln|\psi_\theta|^{2}-\ln|\psi_{\mathrm{Ref}}|^{2}\Bigr]
   \;-\;
   \ln\!\Bigl(
     E_{p_\theta}\bigl[
       |\psi_{\mathrm{Ref}}/\psi_\theta|^{2}\bigr]
   \Bigr).
\end{equation}

The second term still requires the (generally costly and unstable) evaluation of
$N_\theta/N_{\mathrm{Ref}}$.  
%In practice one may either (i) estimate the ratio with a separate Monte-Carlo average, or (ii) adopt the squared-log variant introduced below, which removes the explicit normalization ratio altogether.

A numerically more stable alternative replaces the linear KL integrand in
Eq.~\eqref{eq:kl_def} with the square:

\begin{align}
\label{eq:ll2_def}
\mathcal L_{\rho}
 &= E_{p_\theta}\Bigl[
      \bigl(\ln|\psi_\theta|^{2}-\ln|\psi_{\mathrm{Ref}}|^{2}\bigr)^{2}
    \Bigr] \;=\; E_{p_\theta}[E_L],
\\[3pt]
\text{with}\qquad
E_L &\equiv
      \bigl(\ln|\psi_\theta|^{2}-\ln|\psi_{\mathrm{Ref}}|^{2}\bigr)^{2}.
\end{align}

We borrow the name \emph{local energy} for $E_L$ because its Monte-Carlo
estimate enters the gradient in a way analogous to the local energy in
variational Monte-Carlo optimization.

\subsubsection{Loss gradient calculation}

To derive the formula for the gradient we write the loss as 
\begin{align}
& \mathcal{L}_\rho=	\frac{\int d\boldsymbol{R}\:|\psi_{\theta}(\boldsymbol{R})|^2\left(\ln\left(\frac{|\psi_{\theta}(\boldsymbol{R})|^2}{|\psi_{\text{Ref}}|^{2}}\right)\right)^{2}}{f(\theta)}, \quad f(\theta) = \int d\boldsymbol{R}\:|\psi_{\theta}(\boldsymbol{R})|^{2} \label{supEq:L_rho}
\end{align}

We begin by differentiating the numerator of the loss function with respect to the variational parameters $\theta$. 
\begin{align}
    &\partial_{\theta}\text{Numer} \nonumber=\\ 
    &=\int d\boldsymbol{R}\:\bigg[|\psi_{\theta}(\boldsymbol{R})|^2\left(\ln\left(\frac{|\psi_{\theta}(\boldsymbol{R})|^2}{|\psi_{\text{Ref}}|^{2}}\right)\right)^{2}\partial_{\theta}\ln\left(\psi_{\theta}^{*}(\boldsymbol{R})\right)+|\psi_{\theta}(\boldsymbol{R})|^2\left(\ln\left(\frac{|\psi_{\theta}(\boldsymbol{R})|^2}{|\psi_{\text{Ref}}|^{2}}\right)\right)^{2}\partial_{\theta}\ln\left(\psi_{\theta}(\boldsymbol{R})\right)+\nonumber\\
	&\quad \quad 2|\psi_{\theta}(\boldsymbol{R})|^2\ln\left(\frac{|\psi_{\theta}(\boldsymbol{R})|^2}{|\psi_{\text{Ref}}|^{2}}\right)\left(\partial_{\theta}\ln\left(\psi_{\theta}^{*}(\boldsymbol{R})\right)+\partial_{\theta}\ln\left(\psi_{\theta}(\boldsymbol{R})\right)\right) \bigg]
\end{align}

Dividing by the denominator recasts each integral as a Monte-Carlo expectation with respect to the probability density $p_\theta$, hence every term can now be written explicitly as an average over the density,

\begin{align}
    	&(1) \to \, \, E_{p_\theta}\left[\left(\ln\left(\frac{|\psi_{\theta}|^{2}}{|\psi_{\text{Ref}}|^{2}}\right)\right)^{2}\partial_{\theta}\ln\left(\psi_{\theta}^{*}(\boldsymbol{R})\right)\right]=E_{p_\theta}\left[E_{L}\cdot\partial_{\theta}\left[\log\left(\psi_{\theta}^{*}(\boldsymbol{R})\right)\right]\right] \nonumber\\
	&(2) \to\, \, E_{p_\theta}\left[\left(\ln\left(\frac{|\psi_{\theta}|^{2}}{|\psi_{\text{Ref}}|^{2}}\right)\right)^{2}\partial_{\theta}\ln\left(\psi_{\theta}(\boldsymbol{R})\right)\right]=E_{p_\theta}\left[E_{L}\cdot\partial_{\theta}\left[\log\left(\psi_{\theta}(\boldsymbol{R})\right)\right]\right]  \nonumber\\
	&(3) \to\,  \, E_{p_\theta}\left[2\ln\left(\frac{|\psi_{\theta}|^{2}}{|\psi_{\text{Ref}}|^{2}}\right)\left(\partial_{\theta}\ln\left(\psi_{\theta}^{*}(\boldsymbol{R})\right)+\partial_{\theta}\ln\left(\psi_{\theta}(\boldsymbol{R})\right)\right)\right] = E_{p_\theta}\left[2\sqrt{E_{L}}\left(\partial_{\theta}\left[\log\left(\psi_{\theta}^{*}(\boldsymbol{R})\right)\right]+\partial_{\theta}\left[\log\left(\psi_{\theta}(\boldsymbol{R})\right)\right]\right)\right] \nonumber
\end{align}

Then we notice that 
\begin{align}
    	\log\left(\psi_{\theta}^{*}(\boldsymbol{R})\right)+\log\left(\psi_{\theta}(\boldsymbol{R})\right)=\text{log}\left(|\psi_{\theta}|e^{-i\phi(\theta)}\right)+\text{log}\left(|\psi_{\theta}|e^{i\varphi(\theta)}\right)=2\text{log}\left(|\psi_{\theta}|\right)
\end{align}

This gives, 
\begin{align}
    	\frac{\partial_{\theta}\text{Numer}}{f(\theta)}=E_{p_{\theta}}\left[2\left(E_{L}+2\sqrt{E_{L}}\right)\cdot\partial_{\theta}\left[\text{log}\left(|\psi_{\theta}|\right)\right]\right]
\end{align}

We then carry out a similar analysis for the denominator, 	
\begin{align}
    &\partial_{\theta}\frac{1}{f(\theta)}=-\frac{1}{f\left(\theta\right)}\frac{\partial_{\theta}f(\theta)}{f\left(\theta\right)} \\
    &\frac{\partial_{\theta}f(\theta)}{f\left(\theta\right)}\nonumber =\frac{\int d\boldsymbol{R}\:\left(\partial_{\theta}\psi_{\theta}^{*}(\boldsymbol{R})\right)\psi_{\theta}(\boldsymbol{R})}{\int d\boldsymbol{R}\:|\psi_{\theta}(\boldsymbol{R})|^2}+\frac{\int d\boldsymbol{R}\:\psi_{\theta}^{*}(\boldsymbol{R})\left(\partial_{\theta}\psi_{\theta}(\boldsymbol{R})\right)}{\int d\boldsymbol{R}\:|\psi_{\theta}(\boldsymbol{R})|^2}=\frac{\int d\boldsymbol{R}\:|\psi_{\theta}(\boldsymbol{R})|^{2}\frac{\partial_{\theta}\psi_{\theta}^{*}(\boldsymbol{R})}{\psi_{\theta}^{*}(\boldsymbol{R})}}{\int d\boldsymbol{R}\:|\psi_{\theta}(\boldsymbol{R})|^2}+\frac{\int d\boldsymbol{R}\:|\psi_{\theta}(\boldsymbol{R})|^{2}\frac{\partial_{\theta}\psi_{\theta}(\boldsymbol{R})}{\psi_{\theta}(\boldsymbol{R})}}{\int d\boldsymbol{R}\:|\psi_{\theta}(\boldsymbol{R})|^2}\nonumber\\
	&=E_{p_\theta}\left[\partial_{\theta}\left[\log\left(\psi_{\theta}^{*}(\boldsymbol{R})\right)\right]+\partial_{\theta}\left[\log\left(\psi_{\theta}(\boldsymbol{R})\right)\right]\right]=2E_{p_\theta}\left[\partial_{\theta}\left(\text{log}\left(|\psi_{\theta}|\right)\right)\right]
\end{align}

Combining the differentiated terms produces a compact estimator for the gradient,

\begin{align}
\partial_{\theta}\mathcal{L_\rho}=E_{p_\theta}\left[2\left(E_{L}+2\sqrt{E_{L}}-E_{p_\theta}[E_L] \right)\cdot\partial_{\theta}\left[\text{log}\left(|\psi_{\theta}|\right)\right]\right]    
\end{align}

\subsubsection{Generalization of the gradient result} 

The procedure carries over to any “local-energy’’ functional that depends smoothly on the log-density of the wave function. Let %$\ln|\psi_\theta|^2$.

\begin{align}
\label{eq:ll2_def}
\mathcal{L_{F}} = E_{p_\theta}[E_L], \quad \text{with} \qquad
E_L =
      \mathcal{F}\bigl(\ln|\psi_\theta|^{2}).
\end{align}
where $\mathcal{F}$ is any differentiable scalar function.
The gradient then becomes,
\begin{align}
\partial_{\theta}\mathcal{L_F}=E_{p_\theta}\left[2\left(\mathcal{F}+\mathcal{F}^\prime -E_{p_\theta}[\mathcal{F}] \right)\cdot\partial_{\theta}\left[\text{log}\left(|\psi_{\theta}|\right)\right]\right]    
\end{align}
where $\mathcal{F}^\prime$
  denotes the derivative of $\mathcal{F}$ with respect to its argument

\begin{figure}
    \includegraphics[width=0.98\linewidth]{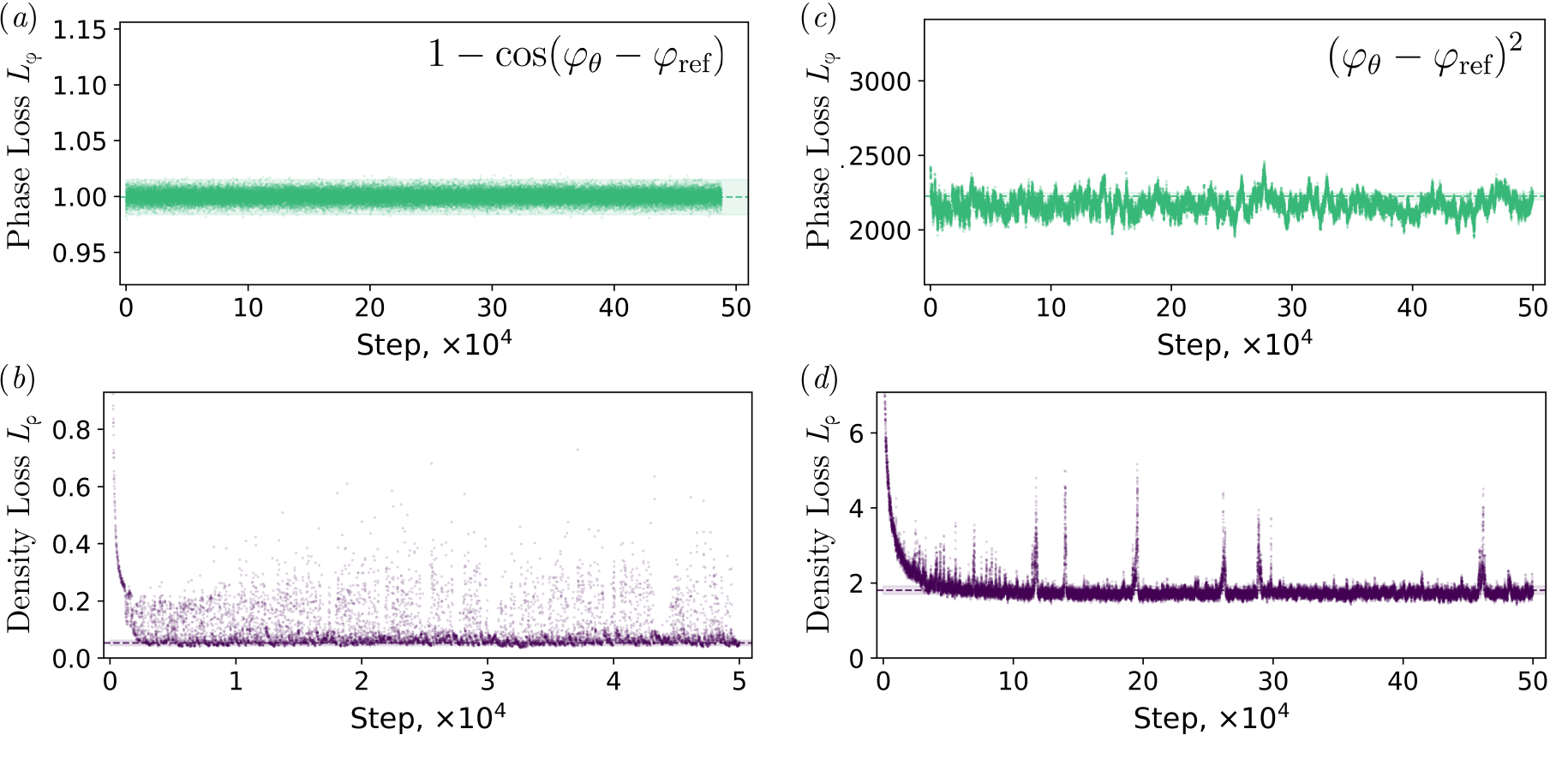}
    \caption{Learning curves for phase $(a, c)$ loss and density loss $(b, d)$ for phase loss functions defined as in eqs. \eqref{supEq:cosPhase} (panels $(a, b)$) and \eqref{supEq:parPhase} (panels $(c, d)$). The network dimensions are: number of layers $= 2$, number of attention heads per layer $= 4$, 
Attention dimension $= 64$ and Perceptron dimension $=256$; with batch size $=2048$, consistent with the dimensions we have used throughout the paper. The phase in both cases is not being learned, lacking sufficient information for optimization. The density loss converges quite efficiently, especially in the case of $\mathcal{L}_{\varphi,2}$ }
    \label{supFig:other_phase_losses}
\end{figure}

\subsection{Current loss}
\subsubsection{Loss calculation}

Minimizing an objective that depends only on the density ensures the variational ansatz reproduces the modulus of the target wave-function, but it says nothing about the \textit{phase}. For systems where topology, circulation, or magnetic fields play a central role—fractional-quantum-Hall droplets, superconductors with quantized vortices, or any state in which transport properties are dictated by Berry phases—capturing the correct phase structure is essential.

The probability current
\begin{align}
    \mathbf{j}(\mathbf{R})=\frac{\hbar}{m} \operatorname{Im}\left[\psi^*(\mathbf{R}) \nabla \psi(\mathbf{R})\right] = \frac{\hbar}{m}|\psi(\mathbf{R})|^2 \nabla \varphi(\mathbf{R})
\end{align}
encodes exactly this missing information. 

By constructing a loss that penalizes the mean-squared difference 
\begin{align}
    \mathcal{L}_j=&\frac{\int d\boldsymbol{R}\:\big(|\psi_{\theta}(\boldsymbol{R})|^{2} \nabla_{\textbf{r}}\varphi_{\theta}-|\psi_{\text{ref}}(\boldsymbol{R})|^{2} \nabla_{\textbf{r}}\varphi_{\text{ref}}\big)^{2}}{\int d\boldsymbol{R}\:|\psi_{\theta}(\boldsymbol{R})|^{2}}
\end{align}
we drive the optimizer to align both the density and the phase gradients. 

The corresponding effective “local energy’’ for the current-matching loss reads
\begin{align}
    E_L = |\psi_{\theta}(\boldsymbol{R})|^{2} \left( \nabla_{\textbf{r}}\varphi_{\theta}-\frac{|\psi_{\text{ref}}(\boldsymbol{R})|^{2}}{|\psi_{\theta}(\boldsymbol{R})|^{2}} \nabla_{\textbf{r}}\varphi_{\text{ref}}\right)^{2}
\end{align}

Computing this expression exactly is numerically delicate: it involves
wave-function ratios outside the log domain. To make the objective practical we introduce two controlled
approximations:

1) \textit{Late-phase activation}.  
We switch on the current-matching term only after the
density-matching loss has converged to high accuracy.
At that stage ${|\psi_{\text{ref}}(\boldsymbol{R})|^{2}} \approx {|\psi_{\theta}(\boldsymbol{R})|^{2}}$ so the troublesome
ratio  is already
close to unity.

2) \textit{Density-independent prefactor}.  We further drop the overall amplitude factor $|\psi_{\theta}(\boldsymbol{R})|^{2}$. The resulting loss still measures the squared
difference between phase gradients and therefore continues to drive the
ansatz toward the correct circulation pattern, while avoiding explicit
amplitude information:

% In practice this local energy is challenging to compute, as it requires working with wave functions and their ratios not in log domain. To mitigate this problem, we slightly alter the loss function. Firstly, we start including the  the current loss in the full loss function when the training with only the density loss has been running long enough and the density pattern has been learning with high enough accuracy. In this case, ${|\psi_{\text{ref}}(\boldsymbol{R})|^{2}} \approx {|\psi_{\theta}(\boldsymbol{R})|^{2}}$. Another simplification that we will do is that we will simply replace the prefactor $|\psi_{\theta}(\boldsymbol{R})|^{2}$ with 1. It is still a valid loss function for phase, as it tries to match the phase gradient of the ansatz with the reference. 

% So, the simplified local energy we use is,

\begin{align}
    E_{L} \to \left(\nabla_{\textbf{r}}\varphi_{\theta}-\nabla_{\textbf{r}}\varphi_{\text{ref}}\right)^{2}
\end{align}

In practice, this simplified local energy retains sensitivity to phase errors,
adds minimal computational overhead (only one extra automatic-differentiation
pass for $\nabla_{\textbf{r}}\varphi_{\theta}$), and sidesteps the numerical
instabilities associated with wave-function ratios in the raw domain.

\subsubsection{Gradient calculation}

Following similar steps as for the density loss gradient, we can  derive the gradient for the current loss, as

\begin{align}
\partial_{\theta}\mathcal{L}_j=E_{p_\theta}\left[2\left(E_{L}-E_{p_\theta}[E_L] \right)\cdot\partial_{\theta}\left[\text{log}\left(|\psi_{\theta}|\right)\right] + 2 \left(\nabla_{\textbf{r}}\varphi_{\theta}-\nabla_{\textbf{r}}\varphi_{\text{ref}}\right) \cdot \partial_\theta (\nabla_\textbf{r} \varphi_\theta )  \right]  
\end{align}

This result can be again generalized to any differentiable function of the phase gradient.

\section{Loss functions for learning the phase}

In this section, we discuss the loss functions designed to directly match the phase of the neural network output, $\varphi_\theta$, to that of the reference function, $\varphi_\text{ref}$. We explored two main types of phase-based loss functions, along with several modifications, but all of them failed for systems with more than 5–6 electrons.

The first loss function is a straightforward mean-square error:
\begin{align}
    \mathcal{L}_{\varphi,1} = &\frac{\int d\boldsymbol{R}|\psi_{\theta}(\boldsymbol{R})|^{2} \:\big(\varphi_{\theta}(\boldsymbol{R})- \varphi_{\text{ref}}(\boldsymbol{R})\big)^{2}}{\int d\boldsymbol{R}\:|\psi_{\theta}(\boldsymbol{R})|^{2}} \label{supEq:parPhase}
\end{align}

The second loss function uses a gauge-invariant quantity, specifically the cosine of the phase difference:
\begin{align}
    \mathcal{L}_{\varphi,2} = &\frac{\int d\boldsymbol{R}|\psi_{\theta}(\boldsymbol{R})|^{2} \:\big(1 - \cos(\varphi_{\theta}(\boldsymbol{R})- \varphi_{\text{ref}}(\boldsymbol{R}))\big)}{\int d\boldsymbol{R}\:|\psi_{\theta}(\boldsymbol{R})|^{2}} \label{supEq:cosPhase}
\end{align}

The gradients of these loss functions are derived in a manner similar to the other loss functions discussed earlier and have been custom-implemented. The density loss is defined in Eq. \eqref{supEq:L_rho}, and the total loss function is taken as the sum of the density and phase losses:     $\mathcal{L} = \mathcal{L}_\rho + \alpha \mathcal{L}_\varphi$,
where $\alpha$ controls the relative weighting of the two components.

Figure \ref{supFig:other_phase_losses} shows the results of simulations for a 9-particle Laughlin state. The simulations clearly indicate that while the density is successfully learned, as evidenced by the decrease of its corresponding loss, the phase loss provides virtually no optimization signal. We experimented with larger architectures and varying $\alpha$ across several orders of magnitude (with $\alpha = 1$ shown here), but the results were consistently the same. This difficulty arises because the phase can vary extremely rapidly in space, making it intrinsically hard to learn.

\section{Training protocol and scaling analysis}

\begin{table}[t]
  \centering
  \caption{Architecture and training hyperparameters}
  \label{tab:train-hparams}
  \begin{tabular}{@{}l@{\hspace{1.5em}}l@{\hspace{4em}}l@{\hspace{1.5em}}l@{}}
    \toprule
    \multicolumn{1}{c}{Hyperparameter} & \multicolumn{1}{c}{Value} &
    \multicolumn{1}{c}{Hyperparameter} & \multicolumn{1}{c}{Value} \\
    \midrule
    %Network Type & Psiformer & Optimizer & KFAC \\
    Number of layers & $1,\,2,\,3, 5$ & Number of determinants & $4,\,8$ \\
    Jastrow factor & None & Optimizer & KFAC\\
    %Head dimensions & $64$ & Layer dimensions & $256$ \\
    KFAC norm constraint & $1\times10^{-3}$ & KFAC damping & $1\times10^{-4}$ \\
    Learning rate & $1\times10^{-3}$ & Batch size & $2048$ \\
    Delay & $1.0\times10^{5}$ & Decay & $1$ \\
    Rescale input & False & Layer norm & True \\
    Precision & FP$32$ & MCMC steps btw iterations & $10$ \\
    \bottomrule
  \end{tabular}
\end{table}

In this section, we describe our training protocol and provide more details of the self-attention architectures considered for the comparative scaling analysis presented in the main text.

Our training protocol is centered around the loss functions $(1)$-$(4)$ presented in the main text, and makes use of a transfer-learning approach to efficiently tackle systems with large particle numbers.
More in detail, our strategy consists of first training the neural network to maximize the overlap with the target wave function for a small system size, and then to leverage the outcome of this problem to tackle larger systems -- the structure of the self-attention architecture presented in Fig.\ $1$ in the main text makes this operation very elegant and economical, as will be discussed below. 
Our systematic transfer-learning approach allows to investigate large system sizes that would be very expensive, if not prohibitive, to tackle otherwise.
\\

$1)$ We begin the training process from a small particle number, such as $N=10$, and train the network by minimizing the loss function $(4)$ with the hyperparameter $\alpha$ increasing gradually from $0$ to $1$ every $2000$ steps following the relation $\alpha(t) = (1 - e^{-20000/t})$, until a sufficiently large number of steps is achieved ($3\times 10^4$ steps guarantee convergence in our case). For our simulations, the learning rate for this part of the training was set to $10^{-3}$.
\\

$2)$ After completion of this first step, we move on to the $N + 1$-particle problem and initialize the network from the weights of the previous $N$-particle training to leverage these converged result.
Because the self-attention layers are independent on the input size, only the parameters for the orbitals and the envelope will be size-mismatched: in this case, the initialization is only partial and a small part of the parameters are initialized from scratch.
\\
Once the network is initialized, the training is performed by minimizing once again the loss function $(4)$, with the important difference that the hyper-parameter is rapidly increased from $0$ to $1$ over the course of $5000$ steps, in order to avoid losing the transferred information from the $N$ particle problem.
Again, the training process is terminated after $3\times 10^4$, as this guaranteed  convergence for the needle problem. For our simulations, the learning rate for this part of the training was set to $10^{-3}$.
\\

$3)$ To further increase the particle number, we repeat the transfer-learning and training protocol presented in $2)$.
\\

%\FG{An iterative transfer learning scheme was recently developed in Ref.\  \cite{Moss_2025} for application of Recurrent Neural Networks on two-dimensional spin lattice systems.}\\

The training protocol described above yields the results shown with the colored circles in Fig.\ $4$ in the main text. The other, “longer”, protocol (square markers in the same figure) simply consists of appending additional $7\times 10^4$ training steps $1)$ and $2)$, where the training now minimizes the fidelity loss function $(1)$ in the main text and. According to our experience, a larger learning rate of $10^{-2}$ works best for this second part of training.
\\ 

All the relevant architectural details, along with the hyper-parameters used for training, are reported in Table \ref{tab:train-hparams}. The different options for the number of layers and number of determinants refer to the three distinct architectures compared in Fig.\ $4$ in the main text.

An important factor in stabilizing training, particularly for larger system sizes, is the normalization of the wave function. Our architecture outputs the real and imaginary components of the wave function, which are subsequently transformed into its logarithmic magnitude and phase. Since we match the logarithmic magnitude of the ansatz wave function to that of the reference, extreme values of the reference magnitude can lead to large variations in the network parameters, resulting in unstable training.

To mitigate this issue, proper normalization of the target wave function is required. In this context, the natural normalization is not the conventional condition $\int d\mathbf{R} \, |\psi_\text{ref}(\mathbf{R})|^2 = 1$, but rather a local scaling such that for each configuration $\{\mathbf{R}\}$ sampled from the reference distribution, the wave function magnitude remains of order unity: $|\psi_\text{ref}(\mathbf{R})| \sim 1$.

To validate the scaling behavior of our computed Laughlin and Moore-Read wave functions, we perform a systematic finite-size scaling analysis of the logarithmic wave function amplitude. Specifically, we compute the median values of $\log |\psi|$ for the reference functions across system sizes ranging from $N = 2$ to $N = 18$ electrons. The median is chosen as a robust statistical measure less sensitive to outliers in Monte Carlo sampling. We fit these median values to the theoretical scaling form $\beta N^2 \log (\alpha N q l_M)$, which arises from the analytical structure of the reference states. Using non-linear least squares fitting via \texttt{scipy.optimize.curve\_fit}, we extract the optimal parameters: $\beta = 0.7951$ and $\alpha = 0.3593$ for the Laughlin state, and $\beta = 0.5609$ and $\alpha = 0.2447$ for the Moore–Read state. The fits show excellent agreement with the numerical scaling, achieving $1 - R^2 \sim 10^{-5}$.

\section{Fractional quantum hall}

\begin{figure}
    \includegraphics[width=0.98\linewidth]{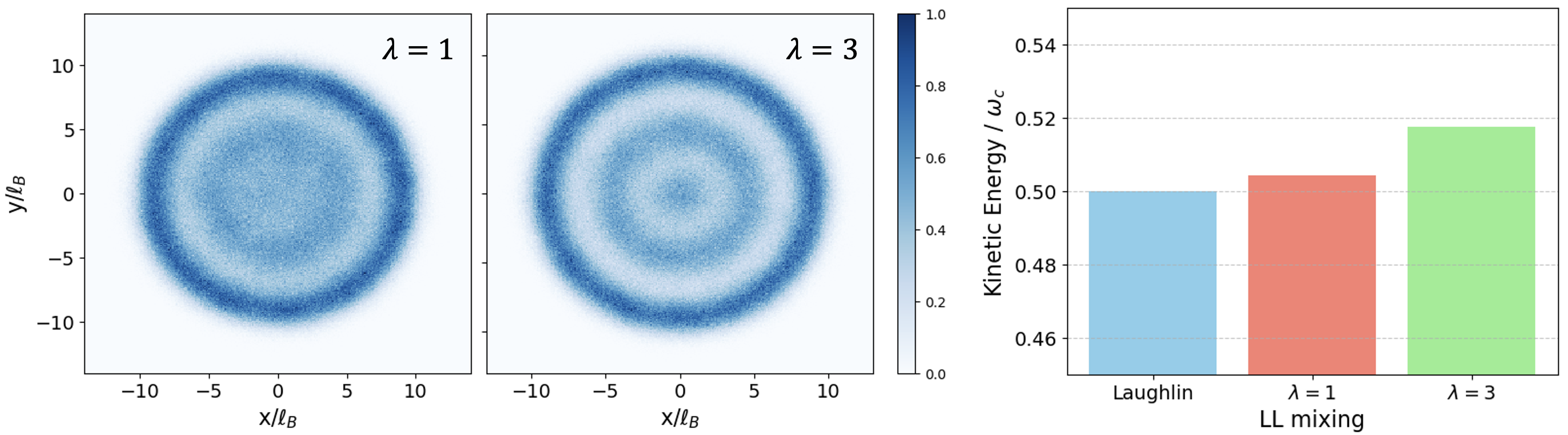}
    \caption{\textbf{LL mixing:} Real space charge density and kinetic energy per particle for 20 electrons with LL mixing parameter $\lambda = 1$ and $\lambda = 3$ }
    \label{fig:ll_mixing}
\end{figure}

Our fractional quantum Hall (FQH) setup consists of $N$ spin-polarized electrons trapped to an infinite 2d plane. Parallel to and at distance $d$ above the plane is a uniformly charged disk of radius $a$ with a total charge $+Ne$, which provides the neutralizing background. To avoid edge reconstruction, we pick $d = 0$, so the positive jellium lies in the same plane as that of electrons. In symmetric gauge, the Hamiltonian of our system can be written as 
\begin{eqnarray}
H  = \sum_{j = 1}^{N} \frac{1}{2}(-i\boldsymbol{\nabla}_j + \frac{1}{2} \boldsymbol{B} \times \mathbf{r}_j )^2 + \sum_{i \neq j} \frac{1}{2\epsilon}\frac{1}{\lvert \mathbf{r}_i - \mathbf{r}_j \lvert} + \sum_{j = 1}^{N} V_c(\mathbf{r}_j) + V_b \,, 
\label{eq:hamiltonian}
\end{eqnarray}
where atomic units, namely $\hbar = e = m_e = 4\pi\epsilon_0 = 1$, are used, $\epsilon$ is the relative dielectric constant, and $V_c$ and $V_b$ are the confining potential and the background self-interaction energy. The background self-interaction is a constant given by $+ 8 N^2/3 \pi \epsilon a$, and the confining potential is given by the following integral expression
\begin{equation}
V_c(\mathbf{r}) = - \frac{N}{\pi a^2} \int_{\lvert \mathbf{r'} \lvert < a} \frac{d\mathbf{r'}^2}{\sqrt{d^2 + \lvert \mathbf{r'} - \mathbf{r} \lvert^2}}
\label{eq:potential}
\end{equation}

\begin{figure}
    \includegraphics[width=0.98\linewidth]{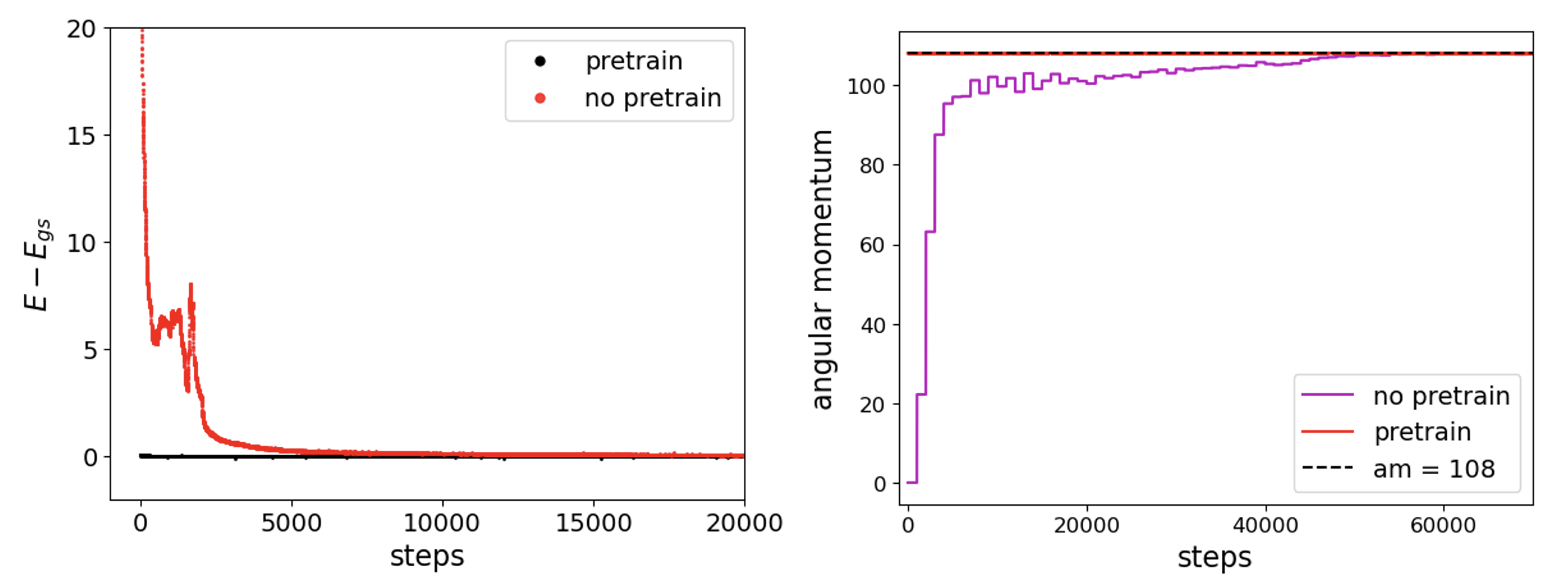}
    \caption{\textbf{Pretrain vs No-pretrain:} Training curves for energy and angular momentum of 9 electrons with $\lambda = 1$.}
    \label{fig:pretrain_vs_nopretrain}
\end{figure}

Further simplifications and efficient implementation are detailed in ~\cite{Teng_2025}. The advantage of our pretrain NN-VMC method is two-fold. Firstly, our neural network solves Schr\"odinger's equation in real space without any truncations, so it captures all Landau levels and the effects of LL mixing, quantified by the dimensionless parameter $\lambda = (e^2/4\pi\epsilon_0 \epsilon \ell_B)/\hbar \omega_c$. To illustrate LL mixing effects, we contrast the charge density profile as well as the kinetic energy per particle for 20 electrons with $\lambda = 1$ and $\lambda = 3$. As shown in Fig.~\ref{fig:ll_mixing}. As we can see, as LL mixing grows stronger, the charge fluctuation will be enhanced, and the contributions from higher Landau levels are no longer negligible. The second advantage of our method comes from the significant speed-up provided by Laughlin pretraining. This is nicely demonstrated in Fig.~\ref{fig:pretrain_vs_nopretrain}, where energy and angular momentum curves are shown for 9 electrons with $\lambda = 1$. A small system size makes both NN-VMC with and without pretrain possible and allows for a direct comparison. As we can see, pretraining to Laughlin state significantly speeds up the convergence and seemingly circumvented the difficulty of reaching the correct angular momentum, arguably the biggest hurdle in the training process.

Regarding computation resources, with pretraining, even 25 electrons with the $3$-layers architecture (our most expensive case, shown in Fig.~3 in the main text) take approximately $72$ hours on one NVIDIA H200 GPU to fully converge (about 50 thousand steps). 

In addition to this, we analyzed the scaling of the energy per particle with the system size. We fit the values with a curve $e_\infty+ c/\sqrt{N}$, where $e_\infty \approx 0.1298$ is the bulk energy per particle in the thermodynamic limit. In the bulk, the system is locally neutral (screening the background), providing the constant term $e_{\infty} N$.
The droplet has a boundary defined by the neutralizing background disk. This leads to an effective surface tension associated with this boundary. Since the radius of the droplet $R$ scales as $R \sim \sqrt{N}$, the perimeter scales as $\sqrt{N}$, and so does the boundary energy. As a result, the correction to the energy per particle scales as ${1}/{\sqrt{N}}$.

For small systems the neural network outputs can be compared to first band projected ED. For $\lambda = 1/3$, the energy per particle (in the unit of $1/\epsilon \ell_B$) for a system of 9 electrons are $1.1346(1)$ and $1.13545$ for NN and ED, respectively \cite{Teng_2025}.

\begin{figure}[h!]
    \includegraphics[width=0.5\linewidth]{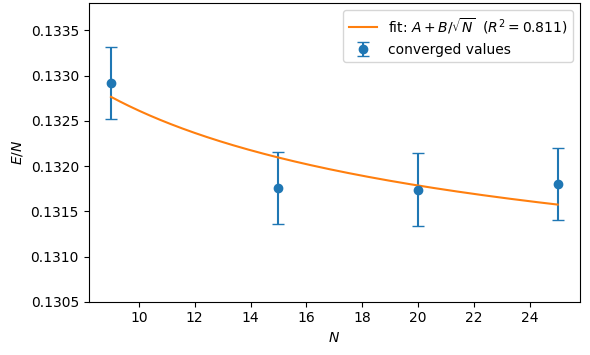}
    \caption{Energy per particle, scaling with the number of particles ($\lambda = 1$). The line represents the fitted curve $e_\infty+ c/\sqrt{N}$.}
\end{figure}

\end{document}